\documentclass[12pt]{article}
\usepackage{epsf}

\newcommand{\vp}{\vec {\it p}}

\newcommand{\be}{\begin{equation}}
\newcommand{\ee}{\end{equation}}
\newcommand{\bmt}{\begin{array}}
\newcommand{\emt}{\end{array}}
\newcommand{\muA}{\mu}
\newcommand{\muB}{\nu}
\newcommand{\muC}{\lambda}
\newcommand{\muD}{\rho}
\newcommand{\seqi}{{\muA}{\muB}{\muC}{\muD}}
\newcommand{\seqv}{{p_1},{p_2},{p_3},{p_4}}

\newcommand{\thm}{\tanh{\left(\frac{m}{2T}\right)}}
\newcommand{\thmN}[1]{\tanh^{#1}{\left(\frac{m}{2T}\right)}}
\newcommand{\ep}[3]{\epsilon^{#1 #2 #3}}
\newcommand{\intn}[1]{\int{\rm d}^{#1} x}
\newcommand{\beN}[1]{\be\label{#1}}
\newcommand{\gammt}{\tilde\Gamma(\tilde a)}

\begin{document}

\title{Parity-violating electromagnetic interactions 
in ${\rm QED_3}$ at finite temperature}
\author{F. T. Brandt$^\dagger$, Ashok Das$^\ddagger$ and J. Frenkel$^\dagger$ 
\\ \\
$^\dagger$Instituto de F\'\i sica,
Universidade de S\~ao Paulo\\
S\~ao Paulo, SP 05315-970, BRAZIL\\
$^\ddagger$Department of Physics and Astronomy,
University of Rochester\\
Rochester, NY 14627-0171, USA}
\maketitle

\bigskip
\noindent

\begin{abstract}
We study the parity-breaking terms generated  by the box diagram in
$2+1$ dimensional thermal QED. These lead, in the long wave
limit, to a gauge invariant extensive action which behaves as $1/T$ at
high temperature. In contrast, the effective action in the static
limit involves leading non-extensive terms proportional to $1/T^3$ at
high temperature, which violate {\it large} gauge invariance. We derive a
non-linear {\it large} gauge Ward identity, which relates the leading 
static terms of different order in perturbation theory and
whose solution coincides with the all order effective action
proposed earlier.
\end{abstract}
\vfill\eject

\section{Introduction:}

It is well known by now that, in odd space-time dimensions, one can
add a topological term to the Lagrangian density of a gauge field, in
addition to the usual Maxwell term. Such a term is known as the
Chern-Simons term \cite{CS} and a theory with such a term is conventionally
called a Chern-Simons theory \cite{dunne:1998lh}. In $2+1$ dimensions,
for example, the Chern-Simons (CS) action has the form \cite{DJT,H}
\begin{equation}
S_{\rm CS} = M \int d^{3}x\,{\rm tr}\, \epsilon^{\mu\nu\lambda}
A_{\mu}\left(\partial_{\nu}A_{\lambda} + {2g\over
3}A_{\nu}A_{\lambda}\right). \label{1}
\end{equation}
Here \lq\lq tr'' denotes trace over the matrix indices of the
gauge fields, $g$ the coupling constant while $M$ is an arbitrary
parameter with the dimensions of mass. 

The CS action has several interesting features. Unlike the standard
Maxwell action for the gauge fields, it is a topological action. In a
theory  with a Maxwell term, the CS action generates a mass
for the gauge fields \cite{DJT}. While it is invariant under {\it small} gauge
transformations, 
the CS action, for a non-Abelian theory, is not invariant under
topologically nontrivial {\it large} gauge transformations. Rather,
its change is proportional to the winding number associated with the
gauge transformation. Explicitly, under
\begin{equation}
A_{\mu} \rightarrow U^{-1}A_{\mu}U + {1\over g}\,U^{-1}\partial_{\mu}
U, \label{2}
\end{equation}
the CS action transforms as
\begin{equation}
S_{\rm CS} \rightarrow S_{\rm CS} + {8\pi^2 M\over g^{2}}\,W, \label{3}
\end{equation}
where
\begin{equation}
W = {1\over 24\pi^{2}} \int d^{3}x\,{\rm Tr}\,\epsilon^{\mu\nu\lambda}
\partial_{\mu}U U^{-1} \partial_{\nu}U U^{-1} \partial_{\lambda}U
U^{-1} \label{4}
\end{equation}
is a topological integer known as the winding number of the gauge
transformation. For vanishing winding number, the gauge
transformations are called {\it small} gauge transformations, while
for any  nontrivial value
of the winding number, they are known as {\it large} gauge
transformations. The CS action clearly is not invariant under {\it
large} gauge transformations. However, the path integral and,
therefore, the theory is, provided the coefficient of the CS term is
quantized as \cite{DJT}
\begin{equation}
{4\pi M\over g^{2}} = n, \label{5}
\end{equation}
with $n$ an integer.

The CS action, in $2+1$ dimensions, is known to violate discrete
symmetries like $P$ and $T$. Furthermore, the mass term for a
fermion (in the irreducible two component representation) is also
known to  violate these symmetries. Therefore, if we
have massive fermions interacting with a background non-Abelian gauge
field, one expects the radiative corrections due to fermions to
generate a  CS term in the effective 
action. In fact, it is known that radiative corrections, at zero
temperature,  shift the value of the tree level CS coefficient
\cite{Redlich} such that (assume, for simplicity, $m>0$)
\begin{equation}
M \rightarrow M - {g^{2}N_{f}\over 8\pi}, \label{6}
\end{equation}
where $N_{f}$ represents the number of fermion flavors. It is clear
now that, even if we start with a consistent theory with tree level
quantization given by Eq. (\ref{5}), the radiative corrections change
this coefficient and the effective theory will continue to be
invariant under {\it large} gauge transformations only for an even
number of fermion flavors. An even number
of fermion flavors is also required to cancel a global anomaly
\cite{Witten}  in such
theories and, therefore, we see that, in such a case, once the tree level CS
coefficient is quantized, the quantum theory continues to have {\it
large} gauge
invariance at the one loop level. In such a theory, it is also known that
there is  no higher loop corrections to the CS coefficient at zero
temperature \cite{coleman} so that the full quantum theory continues
to be invariant under {\it large} gauge transformations. 

In contrast, it was observed that, at finite temperature, the one loop
radiative
corrections due to fermions shift the tree level CS coefficient as
\cite{babu,Zuk,Kao} (We
would see later that this corresponds to a particular limit.)
\begin{equation}
M \rightarrow M - {g^{2}N_{f}\over 8\pi}\,\tanh {\beta m\over 2},
\label{7}
\end{equation}
where $\beta = {1\over T}$ in units of the Boltzmann constant. This,
of course, reduces to Eq. (\ref{6}) when $T\rightarrow 0$. However,
for any nonzero temperature, this is a continuous function and,
therefore, even when the tree level CS coefficient is quantized and
the number of fermion flavors is even, it cannot take a discrete value
as would be required for {\it large} gauge invariance to hold. It would
appear, therefore, that {\it large} gauge invariance would be violated
at finite temperature. On the other hand, this is rather strange
since temperature is not expected to affect gauge invariance, {\it
small} or {\it large}.

The possible understanding of this puzzle has led to a lot of interest
in this topic and only recently, a  mechanism for its resolution has
been found within the context of the $0+1$ dimensional Abelian CS
theory \cite{DLL}. Basically, the resolution of the puzzle in the $0+1$
dimensional model goes as follows. For $N_{f}$ flavors of fermions
interacting with an Abelian gauge background, at zero temperature, the
radiative 
corrections due to fermions generate only the CS term (namely, only
the one point function). On the other hand, at finite temperature, the
effective
action due to fermions can be exactly evaluated and has the form
\begin{equation}
\Gamma_{f} = - i N_{f} \log\left(\cos {a\over 2} + i \tanh {\beta
m\over 2}\,\sin {a\over 2}\right), \label{8}
\end{equation}
where
\begin{equation}
a = \int dt\,A(t), \label{9}
\end{equation}
with $A(t)$ representing the gauge field. This shows that, unlike at
zero temperature, all possible amplitudes are generated in the
effective action at finite temperature. Second, all the terms in the
effective action are non-extensive and, while every individual
term in the effective action violates {\it large} gauge invariance,
for an even number of fermion flavors, the full effective action is
invariant under
\begin{equation}
a \rightarrow a + 2\pi N, \label{10}
\end{equation}
which represents the {\it large} gauge transformation in this case.

By now, the $0+1$ dimensional models have been studied from various
points of view \cite{DD,DD2,BD,DDF}. First of all, since we do not
expect  to be able to
evaluate the effective action in closed form in the $2+1$ dimensional
case, the $0+1$ dimensional theory has been studied exhaustively in
the perturbative approach \cite{DD}. This gives rise to many interesting
features. Similarly, if we were to study the $2+1$ dimensional theory
perturbatively, a signature of {\it large} gauge invariance may lie in
the {\it large} gauge Ward identity. With this in mind, {\it large}
gauge Ward identities have been derived for the $0+1$ dimensional
theories \cite{DDF} which have quite distinctive features. To better
understand
whether the non-extensive structure is special to $0+1$ dimension,
the effective action for a fermion, in $1+1$ dimensions, interacting
with an  Abelian gauge background has also been evaluated at finite
temperature \cite{DDsilva} and it turns out that the effective action,
in this  case,
is extensive although non-local and non-analytic as would be expected
in a thermal background. The analysis of the $0+1$ dimensional model
has also been generalized to $2+1$ dimensional models for a restrictive
gauge background \cite{DGS,FRS}. Namely, it has been shown that for a
single  fermion
interacting with an Abelian gauge background of the form
$A_{0}=A_{0}(t)$ and $\vec{A}=\vec{A}(\vec{x})$, the effective action
has the form \cite{DGS,FRS}
\begin{equation}
\Gamma^\prime = {e\over 2\pi} \int d^{2}x\,\arctan \left(\tanh {\beta
m\over 2} \tan {ea\over 2}\right)\, B, \label{11}
\end{equation}
where the magnetic field is defined to be $B =
\epsilon^{ij}\partial_{i}A_{j}$. 

It is natural to believe that the effective action in Eq. (\ref{11})
does not represent the complete effective action of the fermion theory
in $2+1$ dimensions. In
fact, it is quite clear that the gauge background is quite
restrictive. And, more importantly, the effective action in
Eq. (\ref{11}) does not exhibit non-locality or non-analyticity as
would be expected from a thermal effective action. On the other hand,
it does represent an all order calculation, be it for a very specific
gauge background. It is, of course, quite clear that an exact
evaluation of the effective action in a general gauge background is
impossible. The only way to go beyond the CS action is through
perturbation theory and possibly through the use of the {\it large}
gauge Ward identity. With this in mind, we have decided to evaluate
the parity violating part of the box diagram for a fermion interacting
with an arbitrary Abelian gauge background which may serve as a first
step towards understanding the question of the effective action and,
therefore, {\it large} gauge invariance in the $2+1$ dimensional
theory. Even the calculation of the simple box diagram turns out to be
extremely difficult and we had to make use of symbolic
computer programs in the
intermediate steps. However, the calculation does bring out some
interesting features of the theory. The main results of our analysis
were already reported in \cite{BDF}. In this paper, we describe the
details of our calculation.

The paper is organized as follows. In section {\bf 2}, we compile
our notation as well as various identities in $2+1$ dimensions, which
lead to the fact that all the odd point functions vanish in this
theory. (This is really a consequence of $C$ invariance.)
Consequently, one need to look at only even point functions. In section
{\bf 3}, we exhibit the {\it small} gauge invariance of the fermion
loop at finite temperature. In section {\bf 4}, we discuss the choice
of a {\it small} gauge 
invariant tensor basis which simplifies the calculations. We obtain the
parity violating part of the box diagram at zero temperature as well
as the quartic effective action associated with this. We evaluate
the finite temperature amplitude in  two distinct limits, namely, the
long  wave and the static limits, which
shows that the thermal amplitude is indeed non-analytic. We discuss
various features of the result and construct the corresponding
effective actions. We show that it is really in the static limit that
the question of {\it large} gauge invariance comes up. In section {\bf
5}, we derive a {\it
large} gauge Ward identity and solve for the leading terms in the
static limit, which coincides with the effective action,
Eq. (\ref{11}),  obtained in the
restrictive gauge background. In section {\bf 6}, we present a brief
conclusion along with future directions.

\section{Notations and Conventions:}
      
Let us consider a single flavor of fermion interacting with a
background Abelian gauge field described by the Lagrangian density
\begin{equation}
{\cal L} = \overline{\psi}\left(\gamma^{\mu}(i\partial_{\mu} -
eA_{\mu}) - m\right)\psi. \label{12}
\end{equation}
Here, $e$ represents the electromagnetic coupling strength and we use
a diagonal metric with signatures $(+,-,-)$ as well as assume that
$m>0$. The spinors are two component complex spinors and the Dirac
matrices can be represented in terms of the Pauli matrices $\vec\sigma$
as follows
\begin{equation}
\gamma^{0} = \sigma_{2},\quad \gamma^{1} = i\sigma_{1},\quad
\gamma^{2} = i\sigma_{3}, \label{13}
\end{equation}
so that
\begin{equation}
(\gamma^{0})^{\dagger} = \gamma^{0},\quad (\gamma^{1})^{\dagger} = -
\gamma^{1},\quad (\gamma^{2})^{\dagger} = - \gamma^{2}, \label{14}
\end{equation}
and
\begin{equation}
(\gamma^{0})^{2} = 1 = - (\gamma^{1})^{2} = -
(\gamma^{2})^{2}. \label{15}
\end{equation}

The $2\times 2$ gamma matrices satisfy some interesting relations such as
\begin{equation}
\gamma^{\mu}\gamma^{\nu} = \eta^{\mu\nu} + i \epsilon^{\mu\nu\lambda}
\gamma_{\lambda}, \label{16}
\end{equation}
where $\epsilon^{\mu\nu\lambda}$ represents the totally anti-symmetric
Levi-Civita tensor with $\epsilon^{012}=1$. Relation (\ref{16}) shows
that, unlike in four dimensions, in $2+1$ dimensions, we have
\begin{equation}
{\rm Tr}\;\gamma^{\mu}\gamma^{\nu}\gamma^{\lambda} = 2i
\epsilon^{\mu\nu\lambda}. \label{17}
\end{equation}
It is worth noting here that the gamma matrices satisfy the relation
\begin{equation}
{\rm Tr}\,\gamma^{\mu_{1}}\gamma^{\mu_{2}}\cdots \gamma^{\mu_{2n+1}} =
- {\rm Tr}\, \gamma^{\mu_{2n+1}}\gamma^{\mu_{2n}}\cdots
\gamma^{\mu_{1}}, \label{18}
\end{equation}
which is quite useful in showing that all the odd point functions, in this
theory, vanish which, in turn, is a reflection of charge conjugation
invariance of the theory. (A word of caution here, namely, that this
holds only in the Abelian theory. The presence of internal symmetry
generators invalidates this for non-Abelian theories.) Similarly, for
an even number of gamma matrices, we have
\begin{equation}
{\rm Tr}\,\gamma^{\mu_{1}}\gamma^{\mu_{2}}\cdots \gamma^{\mu_{2n}} =
{\rm Tr}\, \gamma^{\mu_{2n}}\gamma^{\mu_{2n-1}}\cdots
\gamma^{\mu_{1}}, \label{19}
\end{equation}
which helps simplify the calculation of even point functions. There is
one other $2+1$ dimensional identity which is quite useful in
simplifying the calculations, namely, for any arbitrary vector,
$A^{\mu}$, we have
\begin{equation}
A^{\mu}\epsilon^{\nu\lambda\sigma} + A^{\nu}\epsilon^{\lambda\mu\sigma}
+ A^{\lambda}\epsilon^{\mu\nu\sigma} -
A^{\sigma}\epsilon^{\mu\nu\lambda} = 0, \label{20}
\end{equation}
which is really a statement of the fact that in $2+1$ dimensions, we
cannot have a fourth rank anti-symmetric tensor.

\section{Gauge Invariance of the Fermion Loop:}

In trying to evaluate the effective action due to the fermions, let us
next show that, at finite temperature, the $n$-point amplitude
generated by the fermion loop is gauge invariant, at least under {\it
small} gauge transformations.

It is simpler to see the {\it small} gauge invariance in the real time
formalism where there is a doubling of fields \cite{dasbook}. (We will
use  the closed time path formalism \cite{dasbook,Schwinger}.) In this
case,  the propagator acquires a $2\times
2$ matrix structure and an $n$-point amplitude can have both $+$ and
$-$ type of external vertices. For simplicity, we will show gauge
invariance only for amplitudes containing vertices of $+$ type
(namely, the original vertices) although everything can be carried
over to  vertices of other kind.

For the type of amplitude that we are interested in (namely, ones with
$+$ vertices), we need only one component of the fermion propagator, namely,
\begin{equation}
S_{++}(k) = (k\!\!\!\slash + m)\left[{1\over k^{2}-m^{2}+i\epsilon} +
2i\pi\,n_{F}(|k^{0}|) \delta(k^{2}-m^{2})\right]. \label{21}
\end{equation}
Here, $n_{F}$ represents the fermion distribution function. Defining
$q=k'-k$, we have
\vfill\eject
\begin{eqnarray}
S_{++}(k')q\!\!\!\slash S_{++}(k)
& = & \left[{1\over
k'^{2}-m^{2}+i\epsilon} + 2i\pi\,n_{F}(|k'^{0}|)\delta
(k'^{2}-m^{2})\right]\nonumber\\
 & & \times (k\!\!\!\slash ' + m)\left((k\!\!\!\slash '-
m)-(k\!\!\!\slash - m)\right)(k\!\!\!\slash + m)\nonumber\\
 & & \times\left[{1\over
k^{2}-m^{2}+i\epsilon} + 2i\pi n_{F}(|k^{0}|)
\delta(k^{2}-m^{2})\right]\nonumber\\ 
 & = & S_{++}(k) - S_{++}(k'). \label{22}
\end{eqnarray}
This relation is identical to the one at zero temperature.

\begin{figure}[t]
\setlength{\unitlength}{0.005in}%
\begingroup\makeatletter\ifx\SetFigFont\undefined
\def\x#1#2#3#4#5#6#7\relax{\def\x{#1#2#3#4#5#6}}%
\expandafter\x\fmtname xxxxxx\relax \def\y{splain}%
\ifx\x\y   
\gdef\SetFigFont#1#2#3{%
  \ifnum #1<17\tiny\else \ifnum #1<20\small\else
  \ifnum #1<24\normalsize\else \ifnum #1<29\large\else
  \ifnum #1<34\Large\else \ifnum #1<41\LARGE\else
     \huge\fi\fi\fi\fi\fi\fi
  \csname #3\endcsname}%
\else
\gdef\SetFigFont#1#2#3{\begingroup
  \count@#1\relax \ifnum 25<\count@\count@25\fi
  \def\x{\endgroup\@setsize\SetFigFont{#2pt}}%
  \expandafter\x
    \csname \romannumeral\the\count@ pt\expandafter\endcsname
    \csname @\romannumeral\the\count@ pt\endcsname
  \csname #3\endcsname}%
\fi
\fi\endgroup
\begin{picture}(720,300)(120,405)
%
%
\put(538,520){\epsfbox{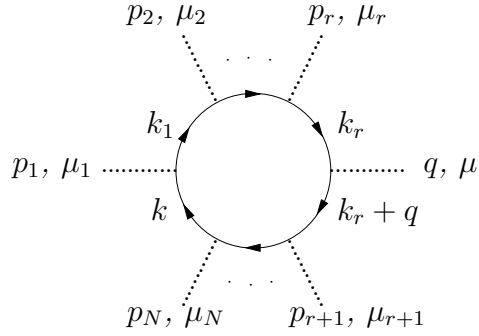}}
\multiput(660,715)(3.0,6.0){12}{\makebox(0.9,1.3){\SetFigFont{10}{12}{rm}.}}
\multiput(582,715)(-3.0,6.0){12}{\makebox(0.9,1.3){\SetFigFont{10}{12}{rm}.}}
\multiput(660,566)(3.0,-6.0){12}{\makebox(0.9,1.3){\SetFigFont{10}{12}{rm}.}}
\multiput(582,566)(-3.0,-6.0){12}{\makebox(0.9,1.3){\SetFigFont{10}{12}{rm}.}}
\multiput(540,640)(-6.7,0.0){12}{\makebox(0.9,1.3){\SetFigFont{10}{12}{rm}.}}
\multiput(704,640)(6.7,0.0){12}{\makebox(0.9,1.3){\SetFigFont{10}{12}{rm}.}}
\put(515,590){\makebox(0,0)[lb]{\smash{\SetFigFont{12}{14.4}{it}$k$}}}
\put(510,680){\makebox(0,0)[lb]{\smash{\SetFigFont{12}{14.4}{it}$k_1$}}}
\put(710,680){\makebox(0,0)[lb]{\smash{\SetFigFont{12}{14.4}{it}$k_r$}}}
\put(710,590){\makebox(0,0)[lb]{\smash{\SetFigFont{12}{14.4}{it}$k_r+q$}}}
\put(370,640){\makebox(0,0)[lb]{\smash{\SetFigFont{12}{14.4}{it}$p_1$,
 ${\mu_1}$}}}
\put(490,800){\makebox(0,0)[lb]{\smash{\SetFigFont{12}{14.4}{it}$p_2$,
 ${\mu_2}$}}}
\put(680,800){\makebox(0,0)[lb]{\smash{\SetFigFont{12}{14.4}{it}$p_r$, 
${\mu_r}$}}}
\put(800,640){\makebox(0,0)[lb]{\smash{\SetFigFont{12}{14.4}{it}$q$, 
 ${\mu}$}}}
\put(660,484){\makebox(0,0)[lb]{\smash{\SetFigFont{12}{14.4}{it}$p_{r+1}$,
 ${\mu_{r+1}}$}}}
\put(490,484){\makebox(0,0)[lb]{\smash{\SetFigFont{12}{14.4}{it}$p_{N}$,
 ${\mu_N}$}}}
%
\end{picture}
\nopagebreak
\caption[f1]{\label{f1}{Fermion loop with $N$ photon lines plus an extra
attached line with momentum $q$ and index $\mu$.
Dotted lines represent photons, and solid lines stand for electrons.}}
\end{figure}

Let us next consider a fermion loop with $N$ photons, carrying momenta
$p_{1},\,p_{2}\cdots\, p_{N}$ and indices $\mu_{1},\mu_{2},\cdots,
\mu_{N}$ in an ordered way and define
\[ k_{r} = k + p_{1} + p_{2} + \cdots + p_{r}, \]
where $k$ represents the momentum in the loop. Let us next attach an
extra photon line with momentum $q$ and index $\mu$ (see Fig.\ref{f1})
between the
photon lines carrying momenta $p_{r}$ and $p_{r+1}$ (all the lines are
of $+$ type).
Contracting this diagram with $q_{\mu}$, we obtain (we
are going to neglect the coupling constants as well as an overall sign
coming from the fermion loop)
\begin{eqnarray}
g_{r}^{\mu_{1}\cdots \mu_{N}}
 & = & \int {d^{3}k\over
(2\pi)^{3}}\,{\rm
Tr}\,S_{++}(k)\gamma^{\mu_{1}}S_{++}(k_{1})\gamma^{\mu_{2}}\cdots
\gamma^{\mu_{r}} \nonumber\\
 &  & \times S_{++}(k_{r}) q\!\!\!\slash
S_{++}(k_{r}+q)\gamma^{\mu_{r+1}}\cdots
S_{++}(k_{N-1}+q)\gamma^{\mu_{N}}\nonumber\\
 & = & \int {d^{3}k\over (2\pi)^{3}}\,{\rm
Tr}\,S_{++}(k)\gamma^{\mu_{1}}S_{++}(k_{1})\cdots
\gamma^{\mu_{r}}\nonumber\\
 & & \times\left(S_{++}(k_{r}) -
S_{++}(k_{r}+q)\right)\gamma^{\mu_{r+1}}\cdots
S_{++}(k_{N-1}+q)\gamma^{\mu_{N}}\nonumber\\
 & = &\!\!\! \int {d^{3}k\over (2\pi)^{3}}\,{\rm
Tr}\,\left[S_{++}(k)\gamma^{\mu_{1}}\cdots
\gamma^{\mu_{r}}S_{++}(k_{r})\gamma^{\mu_{r+1}}\cdots
S_{++}(k_{N-1}+q)\gamma^{\mu_{N}} \right.\nonumber\\
 &  & \left. - S_{++}(k)\gamma^{\mu_{1}}\cdots
\gamma^{\mu_{r}}S_{++}(k_{r}+q)\gamma^{\mu_{r+1}}\cdots
S_{++}(k_{N-1}+q)\gamma^{\mu_{N}}\right]. \label{23}
\end{eqnarray}
Here, we have used the relation Eq. (\ref{22}) in the intermediate
steps.

If we now sum over all the possible insertions of the photon line with
momentum $q$, terms cancel pairwise to give
\begin{eqnarray}
\sum_{r=1}^{N} g_{r}^{\mu_{1}\cdots \mu_{N}} & = & \int {d^{3}k\over
(2\pi)^{3}}\,{\rm
Tr}\,\left[S_{++}(k)\gamma^{\mu_{1}}S_{++}(k_{1})\cdots
S_{++}(k_{N-1})\gamma^{\mu_{N}} \right.\nonumber\\
 &  & \left. - S_{++}(k+q)\gamma^{\mu_{1}}S_{++}(k_{1}+q)\cdots
S_{++}(k_{N-1}+q)\gamma^{\mu_{N}}\right]\nonumber\\
 & = & 0.\label{24}
\end{eqnarray}
Here, we have shifted the variable of integration in the second term
by $k\rightarrow k-q$ to obtain the final result. Such a shift is, of
course, meaningful if the integrand is well behaved. We note here
that, at finite temperature, the temperature dependent terms are
ultraviolet finite and, therefore, such a shift is allowed if the zero
temperature part is well behaved, which we know to be true. This
argument parallels the zero temperature argument \cite{BjD} and shows that the
$n$-point amplitudes generated by the fermion loop are gauge invariant
even at finite temperature. Note that this argument can be easily
extended to any number of dimensions.

As we have said, the gauge invariance of amplitudes with mixed
vertices can be shown in an analogous manner. We simply note here that
with a matrix propagator of the form
\begin{equation}
S(k) = \left(\begin{array}{cc}
              S_{++}(k) & S_{+-}(k)\\
              S_{-+}(k) & S_{--}(k)
              \end{array}\right),\label{25}
\end{equation}
where,
\begin{eqnarray}
S_{+-}(k) & = & 2i\pi (k\!\!\!\slash + m)
(n_{F}(|k^{0}|)-\theta(-k^{0}))\delta(k^{2}-m^{2})\nonumber\\
S_{-+}(k) & = & 2i\pi (k\!\!\!\slash + m)
(n_{F}(|k^{0}|)-\theta(k^{0}))\delta(k^{2}-m^{2})\nonumber\\
S_{--}(k) & = &  (k\!\!\!\slash + m)\left[{1\over k^{2}-m^{2}-i\epsilon} +
2i\pi n_{F}(|k^{0}|) \delta(k^{2}-m^{2})\right], \label{26}
\end{eqnarray}
and a matrix vertex of the form
\begin{equation}
\Gamma^{\mu} = \left(\begin{array}{rr}
                     \gamma^{\mu} & 0\\
                      0 & - \gamma^{\mu}
                     \end{array}\right), \label{27}
\end{equation}
it is easy to see that the generalization of Eq. (\ref{22}) takes the
form \cite{DH1}
\begin{equation}
S(k')(q_{\mu}\Gamma^{\mu})S(k) = S(k) - S(k'). \label{28}
\end{equation}
Furthermore, along with the identities ($q=k'-k$),
\begin{equation}
S_{\pm\mp}(k')q\!\!\!\slash S_{\pm\mp}(k) = 0, \label{29}
\end{equation}
the gauge invariance of any mixed amplitude follows in a completely 
analogous manner.

\section{Box Diagram:}

Since all the odd point functions vanish in this theory and the two
point function is already known (we will come back to the
non-analyticity in the two point function later), the next meaningful
amplitude to evaluate is the four point function. Furthermore, we are
interested only in the parity violating part of this amplitude. This
calculation, of course, is extremely cumbersome. However, as we
have seen in the last section, the four point function has to be invariant
under {\it small} gauge transformations and we expect this to be of
help. While {\it small} gauge invariance
alone predicts uniquely the form of the $n$-point amplitude in the $0+1$
dimensional theory, it is not so in $2+1$ dimensions. For example, we
know that the $n$-point amplitude in $0+1$ dimensions has to be of the
form \cite{DD}
\[ \Pi_{(n)} = \alpha_{n} \delta(p_{1})\delta(p_{2})\cdots
\delta(p_{n-1}), \]
if {\it small} gauge invariance has to hold. However, let us note that
in the $2+1$ dimensional theory, even at the level of the parity
violating four point amplitude, there are several possible structures
that are compatible with {\it small} gauge invariance. Each of the
structures below (and possibly more), for example,
\begin{eqnarray*}
\Pi_{PV}^{\mu\nu\lambda\rho} &\sim &
u^{\mu}u^{\nu}u^{\lambda}\epsilon^{\rho\sigma\tau}
u_{\sigma}p_{4,\tau}\delta(u\cdot p_{1})\delta(u\cdot
p_{2})\delta(u\cdot p_{3}) + perm. \\
\Pi_{PV}^{\mu\nu\lambda\rho} & \sim &
u^{\mu}u^{\nu}\epsilon^{\lambda\rho\tau}p_{4,\tau}\delta(u\cdot
p_{1})\delta(u\cdot p_{2})\delta^{3}(p_{3}+p_{4}) + perm.\\
\Pi_{PV}^{\mu\nu\lambda\rho} & \sim &
u^{\mu}u^{\nu}\epsilon^{\lambda\rho\sigma}u_{\sigma}(u\cdot (p_{3}-p_{4}))
\delta(u\cdot p_{1})\delta(u\cdot
p_{2})\delta(p_{3}^{\perp})\delta(p_{4}^{\perp}) + perm.
\end{eqnarray*}
where $u^{\mu}$ denotes the velocity of the heat bath, is compatible
with {\it small} gauge invariance. This is what makes the calculation hard.
However, one can simplify the calculation somewhat by choosing a ({\it
small}) gauge invariant tensor basis for this amplitude.

\subsection{The Calculation:}

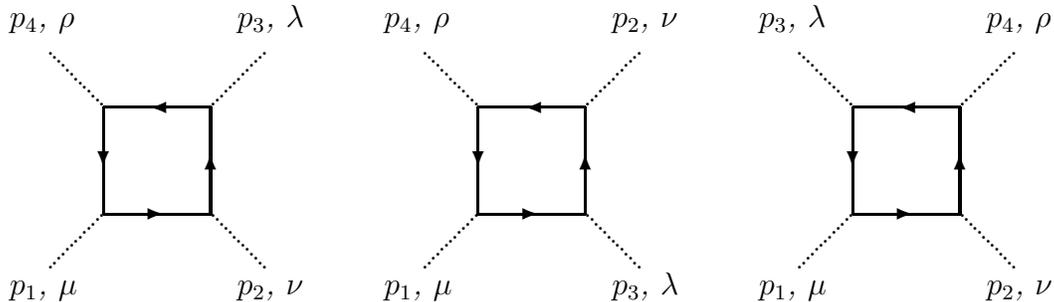
\begin{figure}[t!]
\setlength{\unitlength}{0.007in}%
\begingroup\makeatletter\ifx\SetFigFont\undefined
\def\x#1#2#3#4#5#6#7\relax{\def\x{#1#2#3#4#5#6}}%
\expandafter\x\fmtname xxxxxx\relax \def\y{splain}%
\ifx\x\y   
\gdef\SetFigFont#1#2#3{%
  \ifnum #1<17\tiny\else \ifnum #1<20\small\else
  \ifnum #1<24\normalsize\else \ifnum #1<29\large\else
  \ifnum #1<34\Large\else \ifnum #1<41\LARGE\else
     \huge\fi\fi\fi\fi\fi\fi
  \csname #3\endcsname}%
\else
\gdef\SetFigFont#1#2#3{\begingroup
  \count@#1\relax \ifnum 25<\count@\count@25\fi
  \def\x{\endgroup\@setsize\SetFigFont{#2pt}}%
  \expandafter\x
    \csname \romannumeral\the\count@ pt\expandafter\endcsname
    \csname @\romannumeral\the\count@ pt\endcsname

  \csname #3\endcsname}%
\fi
\fi\endgroup
\begin{picture}(760,232)(60,290)
%
%
\thicklines
\multiput( 80,480)(3.2,-3.2){13}{\makebox(0.7,1.1){\SetFigFont{10}{12}{rm}.}}
\multiput(360,480)(3.2,-3.2){13}{\makebox(0.7,1.1){\SetFigFont{10}{12}{rm}.}}
\multiput(640,480)(3.2,-3.2){13}{\makebox(0.7,1.1){\SetFigFont{10}{12}{rm}.}}
\multiput(240,480)(-3.2,-3.2){13}{\makebox(0.7,1.1){\SetFigFont{10}{12}{rm}.}}
\multiput(520,480)(-3.2,-3.2){13}{\makebox(0.7,1.1){\SetFigFont{10}{12}{rm}.}}
\multiput(800,480)(-3.2,-3.2){13}{\makebox(0.7,1.1){\SetFigFont{10}{12}{rm}.}}
\multiput( 80,320)(3.2,3.2){13}{\makebox(0.7,1.1){\SetFigFont{10}{12}{rm}.}}
\multiput(360,320)(3.2,3.2){13}{\makebox(0.7,1.1){\SetFigFont{10}{12}{rm}.}}
\multiput(640,320)(3.2,3.2){13}{\makebox(0.7,1.1){\SetFigFont{10}{12}{rm}.}}
\multiput(240,320)(-3.2,3.2){13}{\makebox(0.7,1.1){\SetFigFont{10}{12}{rm}.}}
\multiput(520,320)(-3.2,3.2){13}{\makebox(0.7,1.1){\SetFigFont{10}{12}{rm}.}}
\multiput(800,320)(-3.2,3.2){13}{\makebox(0.7,1.1){\SetFigFont{10}{12}{rm}.}}
\put(120,360){\vector( 1, 0){ 45}}
\put(165,360){\line( 1, 0){ 35}}
\put(200,360){\vector( 0, 1){ 45}}
\put(200,405){\line( 0, 1){ 35}}
\put(200,440){\vector(-1, 0){ 45}}
\put(155,440){\line(-1, 0){ 35}}
\put(120,440){\vector( 0,-1){ 45}}
\put(400,360){\vector( 1, 0){ 45}}
\put(445,360){\line( 1, 0){ 35}}
\put(480,360){\vector( 0, 1){ 45}}
\put(480,405){\line( 0, 1){ 35}}
\put(480,440){\vector(-1, 0){ 45}}
\put(435,440){\line(-1, 0){ 35}}
\put(400,440){\vector( 0,-1){ 45}}
\put(680,360){\vector( 1, 0){ 45}}
\put(725,360){\line( 1, 0){ 35}}
\put(760,360){\vector( 0, 1){ 45}}
\put(760,405){\line( 0, 1){ 35}}
\put(760,440){\vector(-1, 0){ 45}}
\put(715,440){\line(-1, 0){ 35}}
\put(680,440){\vector( 0,-1){ 45}}
\put(120,360){\line( 0, 1){ 35}}
\put(400,360){\line( 0, 1){ 35}}
\put(680,360){\line( 0, 1){ 35}}
\put(50,500){\makebox(0,0)[lb]{\smash{\SetFigFont{12}{14.4}{it}$p_4$, 
${\muD}$}}}
\put(330,500){\makebox(0,0)[lb]{\smash{\SetFigFont{12}{14.4}{it}$p_4$, 
${\muD}$}}}
\put(610,500){\makebox(0,0)[lb]{\smash{\SetFigFont{12}{14.4}{it}$p_3$, 
${\muC}$}}}
\put(220,500){\makebox(0,0)[lb]{\smash{\SetFigFont{12}{14.4}{it}$p_3$,
${\muC}$}}}
\put(500,500){\makebox(0,0)[lb]{\smash{\SetFigFont{12}{14.4}{it}$p_2$,
${\muB}$}}}
\put(780,500){\makebox(0,0)[lb]{\smash{\SetFigFont{12}{14.4}{it}$p_4$,
${\muD}$}}}
\put(50,300){\makebox(0,0)[lb]{\smash{\SetFigFont{12}{14.4}{it}$p_1$, 
${\muA}$}}}
\put(330,300){\makebox(0,0)[lb]{\smash{\SetFigFont{12}{14.4}{it}$p_1$, 
${\muA}$}}}
\put(610,300){\makebox(0,0)[lb]{\smash{\SetFigFont{12}{14.4}{it}$p_1$, 
${\muA}$}}}
\put(220,300){\makebox(0,0)[lb]{\smash{\SetFigFont{12}{14.4}{it}$p_2$,
${\muB}$}}}
\put(500,300){\makebox(0,0)[lb]{\smash{\SetFigFont{12}{14.4}{it}$p_3$,
${\muC}$}}}
\put(780,300){\makebox(0,0)[lb]{\smash{\SetFigFont{12}{14.4}{it}$p_2$,
${\muB}$}}}
\end{picture}
\nopagebreak
\bigskip
\caption[f2]{
\label{f2}{
Box diagrams which contribute to the four photon function.}}
\end{figure}

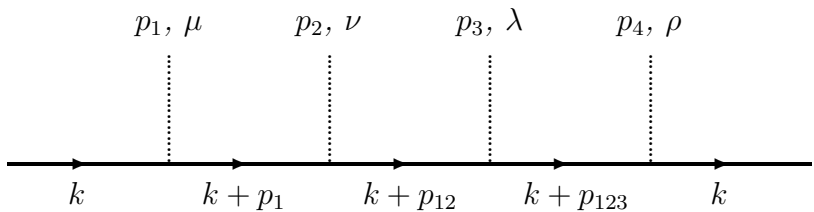
\begin{figure}[t!]
\setlength{\unitlength}{0.007in}%
\begingroup\makeatletter\ifx\SetFigFont\undefined
\def\x#1#2#3#4#5#6#7\relax{\def\x{#1#2#3#4#5#6}}%
\expandafter\x\fmtname xxxxxx\relax \def\y{splain}%
\ifx\x\y   
\gdef\SetFigFont#1#2#3{%
  \ifnum #1<17\tiny\else \ifnum #1<20\small\else
  \ifnum #1<24\normalsize\else \ifnum #1<29\large\else
  \ifnum #1<34\Large\else \ifnum #1<41\LARGE\else
     \huge\fi\fi\fi\fi\fi\fi
  \csname #3\endcsname}%
\else
\gdef\SetFigFont#1#2#3{\begingroup
  \count@#1\relax \ifnum 25<\count@\count@25\fi
  \def\x{\endgroup\@setsize\SetFigFont{#2pt}}%
  \expandafter\x
    \csname \romannumeral\the\count@ pt\expandafter\endcsname
    \csname @\romannumeral\the\count@ pt\endcsname
  \csname #3\endcsname}%
\fi
\fi\endgroup
\begin{picture}(720,204)(120,405)
%
%
\thicklines
\put(220,440){\vector( 1, 0){ 60}}
\put(280,440){\vector( 1, 0){120}}
\put(400,440){\vector( 1, 0){120}}
\put(520,440){\vector( 1, 0){120}}
\put(640,440){\vector( 1, 0){120}}
\put(760,440){\line( 1, 0){60}}
\multiput(340,520)(0.0,-4.1){20}{\makebox(0.9,1.3){\SetFigFont{10}{12}{rm}.}}
\multiput(460,520)(0.0,-4.1){20}{\makebox(0.9,1.3){\SetFigFont{10}{12}{rm}.}}
\multiput(580,520)(0.0,-4.1){20}{\makebox(0.9,1.3){\SetFigFont{10}{12}{rm}.}}
\multiput(700,520)(0.0,-4.1){20}{\makebox(0.9,1.3){\SetFigFont{10}{12}{rm}.}}
\put(315,540){\makebox(0,0)[lb]{\smash{\SetFigFont{12}{14.4}{it}$p_1$,
 ${\muA}$}}}
\put(435,540){\makebox(0,0)[lb]{\smash{\SetFigFont{12}{14.4}{it}$p_2$,
 ${\muB}$}}}
\put(555,540){\makebox(0,0)[lb]{\smash{\SetFigFont{12}{14.4}{it}$p_3$,
 ${\muC}$}}}
\put(675,540){\makebox(0,0)[lb]{\smash{\SetFigFont{12}{14.4}{it}$p_4$,
 ${\muD}$}}}
\put(745,410){\makebox(0,0)[lb]{\smash{\SetFigFont{12}{14.4}{it}$k$}}}
\put(605,410){\makebox(0,0)[lb]{\smash{\SetFigFont{12}{14.4}{it}$k+p_{123}$}}}
\put(485,410){\makebox(0,0)[lb]{\smash{\SetFigFont{12}{14.4}{it}$k+p_{12}$}}}
\put(365,410){\makebox(0,0)[lb]{\smash{\SetFigFont{12}{14.4}{it}$k+p_{1}$}}}
\put(265,410){\makebox(0,0)[lb]{\smash{\SetFigFont{12}{14.4}{it}$k$}}}
\end{picture}
\nopagebreak
\bigskip
\caption[f3]{\label{f3}{One of the four forward scattering amplitudes 
corresponding to the first diagram in \hbox{Fig. 2}.}}
\end{figure}

The graphs which contribute to the four photon amplitude are shown in 
Fig. \ref{f2}. There are three other contributions obtained by
charge conjugation. To evaluate these diagrams, we use
the analytically continued imaginary-time thermal perturbation
theory \cite{dasbook,kapusta,lebellac}. This approach can be
formulated  so as to express the thermal Greens function in
terms of {\it forward scattering} amplitudes \cite{BFT} of an on-shell fermion
in an external electromagnetic field, as depicted in Fig. \ref{f3}.
Each of these {\it forward scattering} amplitude diagrams is obtained by
{\it cutting}  one of the internal lines of the box diagrams in Fig. \ref{f2}.
This, therefore, generates a total of {\it $4\times 6=24$ diagrams},
which can be systematically obtained from the graph in  Fig. \ref{f3}, 
by permutations of the external momenta and polarizations. 
The contribution of the box diagrams, at finite temperature, can then be
written in the form
\begin{equation}
\Pi^{\seqi}(\seqv)=\frac{{ e^4}}{(2\pi)^2} \int
\frac{d^2\vec k}{2\omega_k}\left(n_{F}(\omega_k)-\frac 1 2\right)
\left[
\sum_{ijkl}{ B}^{\mu\nu\lambda\rho}_{(ijkl)}
+ (k\leftrightarrow -k)
\right]. \label{30}
\end{equation}
Here $\omega_k=\sqrt{k^2 +m^2}$, ${n_{F}(\omega_k)}=({{\rm
e}^{\omega_k/T}}+1)^{-1}$, and the sum is over the permutations
$(ijkl)$ of $(1234)$. Each ${ B}$ has a numerator which involves a
trace over the Dirac indices. For example, we have
\beN{31}
\begin{array}{lll}
 &  & { B}^{\mu\nu\lambda\rho}_{(1234)} = \\ & & \\
 &  &\left.\displaystyle{
\frac{{\rm tr}\left[
\left({{\bf \slash}\hskip-.65em\relax k+m}\right)
\gamma^{{\muA}    }
\left({{\bf \slash}\hskip-.65em\relax k+
       {\bf \slash}\hskip-.5em\relax p_{1  }+m}\right)
\gamma^{{\muB}    }
\left({{\bf \slash}\hskip-.65em\relax k+
       {\bf \slash}\hskip-.5em\relax p_{12 }+m}\right)
\gamma^{{\muC} }
\left({{\bf \slash}\hskip-.65em\relax k+
       {\bf \slash}\hskip-.5em\relax p_{123}+m}\right)
\gamma^{{\muD} }\right]}
            {\left({ 2\,k\cdot p_{1  }   +p_{1  }^2}\right)
             \left({ 2\,k\cdot p_{12 }   +p_{12 }^2}\right)
             \left({ 2\,k\cdot p_{123}   +p_{123}^2}\right)}}
\right|_{k_0=\omega_k}, \\
& & 
\end{array}
\ee
where $p_{12} =p_1 +p_2$, etc. 
Here, we are only interested in the contributions from the trace
in Eq. (\ref{31}) which contain odd powers of the mass, since these
will lead to parity-breaking terms (remember that the fermion mass
breaks parity). 

Let us first study the zero temperature contribution coming from the
box diagram, which is
associated with the factor $1/2$ in the first bracket of
Eq. (\ref{30}), as $n_{F}(\omega_k)$ vanishes in this limit. The
computation can be performed explicitly in the {\it low momentum}
region, where $|p_\mu|\ll m$. The result can then be expressed in 
terms of a series in powers of $p/m$, which begins with the leading 
contribution
\begin{eqnarray}
\Pi^{\seqi}_{PV,T=0} & = & -\displaystyle{\frac{{i\,e^4}}{16\pi\,m^6}} 
\left[
\ep{\muA}{\muB}{\alpha}p_1^\alpha\,(p_2)^2 + 
\ep{\muA}{\alpha}{\beta}p_1^\alpha\,p_2^\beta\,p_2^{\muB}
\right] \nonumber\\
& \times & 
\left[
\eta^{\muC\muD}p_3\cdot p_4 - p_3^{\muD} p_4^{\muC}
\right] + {\rm permutations}. \label{32}
\end{eqnarray}
It is interesting to note that this result is consistent with the
Coleman-Hill theorem \cite{coleman}  which implies that, in the
four point Greens function at zero temperature, the terms of order $p$
should be absent. In fact, the above structure shows that the
parity-violating 
contributions, generated by the box diagram at $T=0$, begin only with terms of
order $(p/m)^5$. In the configuration space, the low-energy
effective action associated with Eq. (\ref{32}) can be written in the form
\begin{equation}
\Gamma^4_{PV,T=0} =
-\frac{{e^4}}{64\pi\,m^6}\intn{3}\ep{\mu}{\nu}{\lambda}
F_{\mu\nu}\left(\partial^\tau F_{\tau\lambda}\right)
F^{\rho\sigma} F_{\rho\sigma}, \label{33}
\end{equation}
which is manifestly Lorentz and gauge invariant ({\it small} and {\it
large}).

It is worth pointing out that this is the unique, lowest order (in
derivatives)  parity violating
quartic action that one can construct at zero temperature and can be
thought of as the generalization of the result of Karplus and Neuman
\cite{KN} (to the parity violating amplitude in $2+1$ dimensions). 
Of course, one can naively write down other possible structures, 
for example, of the form
\begin{equation}
S_{PV,T=0}^{4} = \int
d^{3}x\,\epsilon^{\mu\nu\lambda}\partial_{\nu}F_{\lambda\tau}
F_{\mu\rho} F^{\rho\sigma} F_{\sigma}^{\;\tau}. \label{34}
\end{equation}
However, using identities such as in Eq. (\ref{20}) as well as the
Bianchi identity, it is straightforward to show that the two
structures in Eqs. (\ref{33})-(\ref{34}) are related by a simple
multiplicative constant. One can, of course, also construct structures
with  three
epsilon tensors, but they reduce to one of the two forms above. This
shows that the lowest order, parity violating quartic action, at zero
temperature, has a unique form given in Eq. (\ref{33}). In fact, the
identities, in $2+1$ dimensions, are so restrictive that the general
form of the lowest order (in derivatives) parity violating effective
action can be determined to have the form
\begin{equation}
\Gamma_{PV} = \Gamma_{CS} + \sum_{n=1}
a_{n}\epsilon^{\mu\nu\lambda}F_{\mu\nu}
(\partial^{\tau}F_{\tau\lambda}) (F^{\rho\sigma}F_{\rho\sigma})^{n},
\label{35}
\end{equation}
with the coefficient $a_{n}$ to be determined perturbatively ($a_{1}$
is already determined in Eq. (\ref{33})).

The evaluation of the temperature dependent part of the box diagram,
on the other hand, is extremely cumbersome and, as we have mentioned
earlier, we would like to systematize the calculation by first
selecting a gauge invariant basis which we do next.

\subsection{Gauge Invariant Tensor Basis:}

Let us next construct a set of gauge invariant tensor basis for the
parity violating part of the four point amplitude. We note that the
tensors in this basis
must be linear in the Levi-Civita tensor (odd number of epsilons are,
of course, allowed, but reduce to a single epsilon upon using various
identities). Furthermore, the tensor basis should also reflect
symmetry under exchange of external photon lines. At finite
temperature, in addition to the usual tensor structures, we also have
the velocity $u^{\mu}$ of the heat bath and, thus, there are, in
general, many such structures that one can construct. However, it is
practically impossible to carry out the calculation for a general
configuration of momenta. For this reason, we have chosen to work with
a special configuration of momenta, namely,
\begin{equation}
p_{1} = p_{2} = p_{3} = p = - {1\over 3} p_{4}. \label{36}
\end{equation}
In this special configuration, the number of linearly independent,
gauge invariant tensor structures is rather easy to determine. For
example, for tensor structures where the Levi-Civita tensor has two
free indices, there are only two linearly independent structures
possible, namely,
\begin{eqnarray}
T_{1}^{\mu\nu\lambda\rho} & = &
\epsilon^{\sigma\lambda\rho}p_{\sigma}\left(\eta^{\mu\nu} -
{p^{\mu}p^{\nu}\over p^{2}}\right) + perm.\nonumber\\
T_{2}^{\mu\nu\lambda\rho} & = &
\epsilon^{\sigma\lambda\rho}p_{\sigma}\left(u^{\mu} - {p\cdot u\over
p^{2}}p^{\mu}\right)\left(u^{\nu} - {p\cdot u\over
p^{2}}p^{\nu}\right) + perm. \label{37}
\end{eqnarray}
These two independent structures are, in fact, quite easy to
understand intuitively. Let us recall that, for the parity conserving
part of the two point function, there are two independent tensor
structures at finite temperature \cite{dasbook,Weldon} (there are
really three structures 
with a constraint) and the parity violating part of the self-energy
has a unique structure with the epsilon tensor. The two structures
above simply arise as products of the parity violating structure with
the two independent parity conserving structures.

One can similarly look for tensor structures where two of the indices
of the Levi-Civita tensor are contracted. There are again only two
linearly independent, gauge invariant tensor structures of this kind
that one can construct and they have the forms
\begin{eqnarray}
T_{3}^{\mu\nu\lambda\rho} & = &
\epsilon^{\sigma\tau\rho}u_{\sigma}p_{\tau}\left(u^{\lambda} - {p\cdot
u\over p^{2}}p^{\lambda}\right)\left(u^{\mu} - {p\cdot u\over
p^{2}}p^{\mu}\right)\left(u^{\nu} - {p\cdot u\over
p^{2}}p^{\nu}\right) + perm.\nonumber\\
T_{4}^{\mu\nu\lambda\rho} & = &
\epsilon^{\sigma\tau\rho}u_{\sigma}p_{\tau}\left(u^{\lambda} - {p\cdot
u\over p^{2}}p^{\lambda}\right)\left(\eta^{\mu\nu} - {p^{\mu}p^{\nu}\over
p^{2}}\right) + perm. \label{38}
\end{eqnarray}
The four structures in Eqs. (\ref{37})-(\ref{38}) represent a complete
set of linearly independent, gauge invariant basis for the parity
violating part of the box diagram in this special momentum
configuration. (There are other structures possible, but they are not
linearly independent.) Therefore, the parity violating part of the
four point amplitude can be written as
\begin{equation}
\Pi_{(4),PV}^{\mu\nu\lambda\rho} = \sum_{i=1}^{4} C_{i}
T_{i}^{\mu\nu\lambda\rho}. \label{39}
\end{equation}
where the coefficients $C_{i}$ are to be determined from the actual
evaluation of the Feynman diagrams. Explicit calculation shows that
$C_{3}=C_{4}=0$, so that the parity violating part of the four point
amplitude can be expressed in terms of only the first two structures
in Eq. (\ref{37}).

\subsection{Non-analyticity:}

In evaluating the box amplitude at finite temperature, one faces yet
another difficulty. Namely, thermal amplitudes are known to be
non-analytic at the origin in the energy-momentum plane \cite{WBD}, which is
understood as resulting from new branch cuts that develop at finite
temperature due to the possibility of additional channels of
reaction. This also translates to the fact that the temperature
dependent effective action has a non-analytic structure \cite{DH2}. This is
not of importance in $0+1$ dimension where there is no non-analyticity. But,
this becomes quite crucial in higher dimensions. Thus, for example,
the parity violating part of the self-energy, in $2+1$ dimensions, has
the form
\begin{equation}
\Pi_{PV}^{\mu\nu}(p) = -{ime^{2}\over
(2\pi)^{2}}\,\epsilon^{\mu\nu\lambda} p_{\lambda} \int {d^{2}k\over
\omega_{k}}\,\tanh\left({\omega_{k}\over 2T}\right)\left({1\over
p^{2}+2k\cdot p} + k\leftrightarrow -k\right). \label{40}
\end{equation}
Although it is not widely appreciated, the integrand in Eq. (\ref{40})
is non-analytic at $p^{\mu}=0$ and depending on how one evaluates the
integral, the result would be different. For example, in the long wave
limit (LW), the leading term, at high temperature, of the parity
violating part of the self-energy takes the form
\begin{equation}
\Pi_{PV}^{\mu\nu (LW)}(p^{0},\vec{p}=0) = -{ie^{2}\over
8\pi}\,{m\over T}\ln\left({m\over T}\right)\,\epsilon^{0\mu\nu}p^0 + \cdots,
\label{41}
\end{equation}
giving rise to a leading quadratic effective action of the form
\begin{equation}
\Gamma_{CS}^{(LW)} = {e^{2}m\over 16\pi T}\,\ln {m\over T} \int
d^{3}x\,\epsilon^{0ij}\,A_{i}E_{j}. \label{42}
\end{equation}

In contrast, in the static limit (S), the leading behavior of the
parity violating term in the self-energy has the form
\beN{43}
\Pi_{PV}^{\mu\nu (S)} (p^{0}=0,\vec{p}) =  {ie^{2}\over
4\pi}\,\thm \epsilon^{\mu\nu j} p_{j} + \cdots
\ee
giving rise to a leading quadratic effective action of the form
\begin{equation}
\Gamma_{CS}^{(S)} = {e^{2}\over 4\pi}\thm\int d^{3}x\, A_{0}
B. \label{44}
\end{equation}
There are several things to note from this analysis. First, the form of the
effective actions, in the two different limits, are quite
different. Their leading temperature dependence is also quite
distinct. And, finally, if we think of finite temperature as
compactifying the time direction and, thereby, inducing a {\it
large} gauge transformation, then, the effective action in the long wave
limit is invariant under such {\it large} gauge transformations while
the problem of {\it large} gauge invariance really manifests in the
static limit. Without going into detail, we would like to point out
here that the leading contribution to the CS term vanishes if we
approach the origin $p^{\mu}=0$ along a light-like direction. It is
also worth pointing out here that the original calculation of the CS
term \cite{babu} corresponds to evaluating it in the static limit (the
question of non-analyticity was not very much appreciated then).

It is clear, therefore, that in evaluating the box diagram, at finite
temperature, we expect the amplitude to be non-analytic as well. In fact, as
in the case of the self-energy, we are going to evaluate this
amplitude only in the long wave and the static limits. Let us first
concentrate on the long wave limit. In this limit 
($\vec{p}=0$) we have the relation $p^{\mu}=(p\cdot u)u^{\mu}$ and,
therefore,  it is clear that three of the four basis tensor
structures in Eqs. (\ref{37})-(\ref{38}) identically vanish (as we
have mentioned earlier, the last two structures do not contribute to
the parity violating part of the four point amplitude at all), namely,
\[
T_{2}^{\mu\nu\lambda\rho} = 0 = T_{3}^{\mu\nu\lambda\rho} =
T_{4}^{\mu\nu\lambda\rho},
\]
and the only non-vanishing basis tensor takes the simple form (with a
multiplicative factor taken out)
\begin{equation}
T_{1}^{\mu\nu\lambda\rho} = \epsilon^{\sigma\lambda\rho}u_{\sigma}
(\eta^{\mu\nu} - u^{\mu}u^{\nu}) + perm. \label{45}
\end{equation}
In this case, the amplitude can be written as
\begin{equation}
\Pi_{PV}^{\mu\nu\lambda\rho (LW)} = C_{1}
\epsilon^{\sigma\lambda\rho}u_{\sigma} (\eta^{\mu\nu} -
u^{\mu}u^{\nu}) + perm. \label{46}
\end{equation} 
We note here from the explicit structure in Eq. (\ref{46}) that in the
long wave limit, the amplitude is nontrivial only when all the external
indices take spatial values. Furthermore, the coefficients $C_{1}$
which depend on the momenta, temperature etc,
are to be evaluated from the Feynman diagram and have the form
\begin{equation}
C_{1} = -{6ie^{4}\,m\,p_0\over \pi} \int_{0}^{\infty}
d|\vec{k}|\,{|\vec{k}|\over \omega_{k}} \tanh {\omega_{k}\over 2T}
{(3\omega_{k}^{2}-5m^{2}+2p_0^{2})\over
(p_0^{2}-\omega_{k}^{2})(p_0^{2}-4\omega_{k}^{2})
(9p_0^{2}-4\omega_{k}^{2})}. \label{47}
\end{equation}
For $|p_{0}|\ll m$, we can expand this in a series of the form
\begin{equation}
C_{1} =\frac{i\,e^4\,m\,T\,p_0}{16}\,\sum_{l=-\infty}^{\infty}
\left\{
\left(\frac{5\,m^2}{\Delta_l^6}+\frac{3}{\Delta_l^4}\right)
\ln{\left(1+\frac{\Delta_l^2}{m^2}\right)}
-\frac{5}{\Delta_l^4} -\frac{1}{2\,m^2\Delta_l^2}
\right\}, \label{48}
\end{equation}
where $\Delta_l \equiv (2l+1)\,\pi\,T$. In the high temperature
limit, the leading contribution comes from the last term in
Eq. (\ref{48}). Performing the summation over $l$, we then obtain that
\begin{equation}
C_{1}(T\gg m) = - \frac{i\,e^4\,p_0}{128}\frac{1}{m\,T}. \label{49}
\end{equation}
Therefore, we see that the leading contribution, in the long wave
limit, comes from an extensive effective action of the form 
\begin{equation}
\tilde\Gamma_{PV}^{4 (LW)} = \frac{e^4}{512\,m\,T}
\intn{3}\,\epsilon^{0ij}
E_i\left(\partial_t^{-1}E_j\right)
\left(\partial_t^{-1}E_k\right)
\left(\partial_t^{-1}E_k\right), \label{50}
\end{equation}
where $\vec E$ denotes the electric field. This action is non-local and
manifestly gauge invariant (both under {\it small} and {\it large} gauge
transformations) much like the quadratic effective action in the long
wave limit. We would like to note here that, in the long wave limit,
we have evaluated the amplitude for arbitrary values of the energies,
but have chosen to present the results only for the special
configuration of Eq. (\ref{36}) for simplicity.

Next, let us turn to the discussion of the thermal behavior of the box
diagram in the static
limit, where $p_{i}^{0}=0$. In this case, due to the very complicated
angular  integrations,
the calculations are extremely difficult, even when using 
{\it computer algebra}. As a result, we
have restricted ourselves, in this calculation, to the special
configuration of the external spatial
momenta (of Eq. (\ref{36})), where $\vec p_1=\vec p_2=\vec p_3=\vec p=
-{1\over 3}\vec p_4$. In this case, we note that the tensor structure
$T_{2}^{\mu\nu\lambda\rho}$ would  give contributions only to
$\Pi_{PV}^{000i}$ whereas $T_{1}^{\mu\nu\lambda\rho}$ would give
contributions to both $\Pi_{PV}^{000i}$ as well as
$\Pi_{PV}^{0ijk}$ (in the rest frame of the heat bath). Thus, we see
that, in  the static limit, the
amplitude can have only an odd number of temporal indices (unlike the
long wave limit). Let us parameterize the two nontrivial amplitudes as
\begin{eqnarray}
\Pi_{PV}^{000i (S)} & = & 2(C_{1}+C_{2}) \epsilon^{0ij}
\frac{p_{j}}{|\vp|} =
-\frac{1}{4\,|\vp|^2} \epsilon^{0\;ij} p_{j} \Pi_1(\vp,T)\label{51}\\
\Pi_{PV}^{0 ijk (S)} & = & 2C_{1} \epsilon^{0kl}
\frac{p_{l}}{|\vp|} 
\left({p^{i}p^{j}\over \vec{p}^{2}} + \eta^{ij}\right)
\nonumber\\
 & = &  \frac{1}{12\,|\vp|^4}
\epsilon^{0kl}p_{l} 
\left(|\vec p|^2 \eta^{ij} + p^{i}p^{j}\right) \Pi_2(\vp,T), \label{52}
\end{eqnarray}
where $\Pi_{1,2}(\vp,T)$ are rather complicated functions of the momenta
and the temperature. However, for small momenta, namely, $|\vp|\ll
m,T$, we can expand these in a powers series in the momenta and each
term in the series can be evaluated in a straightforward
manner. Thus, for example, the leading
term in $\Pi_1$, in this domain, can be evaluated to have the form
\begin{equation}
\Pi_1(\vp,T) =
\frac{6\,i\,e^4}{4 \pi}\left[\thm - \thmN{3}\right]\frac{|\vp|^2}{T^2}
+ O\left(\frac{|\vp|^4}{m^2\,T^2}\right). \label{53}
\end{equation}
In the high temperature limit, this behaves as ${1\over T^3}$,
which is quite different from the leading ${1\over T}$ behavior of 
the result (\ref{49}) in the long wave limit. Let us note
here that $\Pi_{2}$ can also be evaluated in a similar fashion and has
the leading high temperature behavior
\begin{equation}
\Pi_{2}(\vp,T) = - {17ie^{4}\over 16800\pi}\,{m^{3}\over
T^{7}}\,p^{4}. \label{54}
\end{equation}
It is interesting that terms with lower powers of momentum in
$\Pi_{2}$ identically vanish. As a result, we see that the leading
term in $\Pi_{PV}^{0ijk (S)}$ is strongly suppressed at high
temperature compared with $\Pi_{PV}^{000i (S)}$ which, in turn, is
suppressed relative to $\Pi_{PV}^{ijkl (LW)}$.

The leading contribution given in Eqs. (\ref{51}) and (\ref{53})
can be associated with the effective non-extensive action (remember
that $p_{4}=-3p$) 
\be
\tilde\Gamma^{4 (S)}_{PV}=\frac{e^4\,T}{48 \pi}
\left[\thm - \thmN{3}\right]\intn{3}\,a_0^3\,B,\label{55}
\ee
where we have defined
\be
a_0 = \int_0^{\beta}{\rm d}t\, A_0(t,\vec x)
\ee
and $B$ is the magnetic field.
This form, which may also hold in the quasi-static limit,
is consistent with the result derived from the all-orders
effective action noted earlier in the special gauge background (see
Eq. (\ref{11})). We note here that the effective action
that would give rise to the amplitude $\Pi_{PV}^{0ijk (S)}$ in
Eqs. (\ref{52}) and (\ref{54}) can also be determined in a similar
manner, but is highly suppressed at high temperature and, unlike the
non-extensive action in Eq. (\ref{55}), would have an extensive, be it
non-local structure characteristic of thermal actions. Let us note
here that our calculations have been done in the small momentum
approximation (which would correspond to a derivative expansion of the
effective action). It is well known that \cite{BD}, in such an expansion, it is
impossible to pick out delta function structures characteristic of
non-extensive actions unless one sums the series, which in the present
case is simply impossible. In fact, even the evaluation of the leading
term in the small momentum expansion already pushes us to the limit of
our computational abilities (we really mean even with the use of
computers). Therefore, in isolating delta function structures, we have
been guided by our earlier experience from the studies in $0+1$
dimension \cite{DD}, namely, that if an amplitude has a delta function
structure, then, in the small
momentum expansion, the amplitude vanishes if the variable has a
nonzero value and is nonzero only when the variable assumes a
vanishing value. This we have checked explicitly. It remains an open
question as to whether one can find a better way of isolating delta
function structures from a calculation of the leading term in the
small momentum expansion.

To conclude this section, therefore, we have found that the
temperature dependent part of the parity violating four point
amplitude is non-analytic, much like the self-energy. The effective
actions, in general, contain both extensive as well as non-extensive
terms. In the long wave limit, the leading term at high temperature
goes as ${1\over T}$ and the effective action associated with this is
extensive. Furthermore, this action is invariant under both {\it
small} and {\it large} gauge transformations, much like the CS action
in the long wave limit. In the static limit, the leading term in the
effective action is non-extensive and behaves as ${1\over T^{3}}$ at
high temperature. Furthermore, while this action is invariant under
{\it small} gauge transformations, it violates {\it large} gauge
invariance. Thus, {\it large} gauge invariance seems to hold order by
order in the long wave limit, while it is the static limit where {\it
large} gauge invariance appears to be an issue at every order.

\section{{\it Large} Gauge Ward Identity:} 

To a given order, the quasi-static perturbative contributions are not
invariant under {\it large} gauge
transformations generated by $e\,a_{0}\rightarrow e\,a_{0} + 2\pi\,N$,
where $N$ is
a topological integer. But one can derive, in this case, a Ward
identity for {\it large} gauge invariance, which relates the amplitudes
obtained in perturbation theory. To this end, motivated by the 
structure of Eq. (\ref{55}), let us write the 
corresponding all order effective action in the form
\beN{56}
\tilde\Gamma^{(S)}=\frac{e\,T}{2 \pi}\intn{3}\,\gammt B,
\ee
where $\tilde a = e\,a_0$. It has been noted in \cite{DD} that in
the special background $A_0=A_0(t)$ and $\vec A=\vec A(\vec x)$,
$\gammt $ corresponds to the real part of the
effective action $\Gamma_f$ in Eq. (\ref{8}), with the identification
$a\rightarrow\tilde a$.
This action obeys, for a single fermion flavor, the 
{\it large} gauge Ward identity \cite{DDF}
\beN{57}
\frac{\partial^2\Gamma^{(1)}}{\partial\tilde a^2}=
i\left[\frac 1 4 - 
\left(\frac{\partial\Gamma^{(1)}}{\partial\tilde a}\right)^2\right],
\ee
where the one point function has the value
\beN{58}
\left.\frac{\partial\Gamma^{(1)}}{\partial\tilde a}
\right|_{\tilde a = 0} =
{1\over 2} \tanh {\beta m\over 2}.
\ee

In order to derive the {\it large} gauge Ward identity satisfied by
$\gammt =\Re\left[\Gamma^{(1)}(\tilde a)\right]$,
we write
\beN{59}
\Gamma^{(1)}(\tilde a)=\gammt  + i\,I(\tilde a),
\ee
where $I$ denotes the imaginary part of the action $\Gamma^{(1)}$,
and substitute this relation into the nonlinear equation
(\ref{57}). Equating to zero the resulting real and imaginary parts, we
obtain the following system of coupled equations
\beN{60}
\frac{\partial^2\tilde\Gamma}{\partial\tilde a^2}=
2\,\frac{\partial\tilde\Gamma}{\partial\tilde a}\,
\frac{\partial I}{\partial\tilde a};\;\;\;\;\;\;\;\;
\frac{\partial^2 I}{\partial\tilde a^2}=\frac 1 4 +
\left(\frac{\partial I}{\partial\tilde a}\right)^2-
\left(\frac{\partial\tilde\Gamma}{\partial\tilde a}\right)^2 .
\ee
We must now eliminate from the first equation 
${\partial I}/{\partial\tilde a}$, so as to express 
${\partial^2\tilde\Gamma}/{\partial\tilde a^2}$ solely in terms of
functionals of $\tilde\Gamma$. After some analysis, it turns out that
a consistent solution of the above set of equations requires
${\partial I}/{\partial\tilde a}$ to have the form
\beN{61}
\frac{\partial I}{\partial\tilde a} = 
A\,\sin(\omega \tilde\Gamma) + B\,\cos(\omega \tilde\Gamma),
\ee
where the coefficients $A$ and $B$, as well as the frequency $\omega$,
must be determined from the boundary conditions. One of these
conditions can be read directly from (\ref{58}) and the fact that
$\gammt =\Re\left[\Gamma^{(1)}(\tilde a)\right]$. The
other condition follows from the form (\ref{56}) of the effective
action $\tilde{\Gamma}^{(S)}$ which, as a consequence of invariance
under charge conjugation, is a functional involving only even powers of
$A_\mu$. Consequently, $\gammt$ must contain only odd powers of
$\tilde a$ and therefore, in particular,
${\partial^2\tilde\Gamma}/{\partial\tilde a^2}$
should vanish at $\tilde a = 0$. These conditions, together with the
set of Eqs. (\ref{60}), determine uniquely $A$, $B$ and $\omega$ in 
Eq. (\ref{61}), so that
\beN{62}
\frac{\partial I}{\partial\tilde a}=
\frac{1}{2\,\sinh \beta m}\,\sin(2\,\tilde\Gamma).
\ee

Using this form, we find from the first relation in Eq. (\ref{60})
that $\gammt$ satisfies the {\it large} gauge Ward identity
\beN{63}
\frac{\partial^2\tilde \Gamma}{\partial\tilde a^2}=
\frac{1}{{\rm sinh}\beta m} 
\frac{\partial\tilde \Gamma}{\partial\tilde a}\,
\sin{\left(2\tilde\Gamma\right)}.
\ee
This identity, which  reflects the underlying 
{\it large} gauge invariance of the quasi-static ${\rm QED_3}$
theory, relates higher point Greens functions to lower ones. However,
unlike the Ward identity for {\it small} gauge transformations, the relation
(\ref{63}) is nonlinear. In some sense, this is expected for {\it large}
gauge transformations which are topologically
nontrivial. The relation
(\ref{63}), in fact, allows us to check for {\it large} gauge
invariance perturbatively. Note from Eqs. (\ref{44}), (\ref{55}) and
(\ref{56}) that
\begin{equation}
\left.{\partial\tilde{\Gamma}\over \partial \tilde{a}}\right|_{\tilde{a}=0} =
{1\over 2}\tanh {\beta m\over 2},\qquad \left.{\partial^{3}
\tilde{\Gamma}\over \partial \tilde{a}^{3}}\right|_{\tilde{a}=0} = {1\over
4} \left(\tanh {\beta m\over 2} - \tanh^{3} {\beta m\over
2}\right).\label{64}
\end{equation}
The identity in Eq. (\ref{63}) leads to  (remember that
$\tilde{\Gamma}$ is odd in $\tilde{a}$ and hence vanishes for
$\tilde{a}=0$),
\begin{equation}
\left.{\partial^{3}\tilde{\Gamma}\over \partial
\tilde{a}^{3}}\right|_{\tilde{a}=0} = {2\over \sinh \beta
m}\,\left(\left.{\partial\tilde{\Gamma}\over \partial
\tilde{a}}\right|_{\tilde{a}=0}\right)^{2}.\label{65}
\end{equation}
This can be easily seen to hold from the relations in
Eq. (\ref{64}). In fact, the solution of the Ward identity (\ref{63}),
subject to the above boundary conditions, is given by
\beN{66}
\gammt =
{\rm arctan}\left[\tanh {\beta m\over 2}\tan\left(\frac{\tilde
a}{2}\right)\right].
\ee
Note that in this solution, which sums up the leading perturbative
effects in this region, the tangent is invariant under the {\it large} 
gauge transformations $\tilde a\rightarrow\tilde a + 2\pi\,N$.
 (Incidentally, {\it large} gauge invariance would also
require quantization of the magnetic flux, which we do not get into
here.) 
Substituting the form (\ref{66}) in the expression (\ref{56}), 
we obtain for $\tilde\Gamma(s)$ a result which agrees, 
in the static limit of  ${\rm QED_3}$,
with the parity-breaking effective action previously discussed in the
literature \cite{DGS,FRS}.

\section{Conclusion:}

In this paper, we have studied the radiatively generated parity
violating part  of the four point amplitude in a theory of a single
fermion interacting with an arbitrary Abelian gauge background in
$2+1$ dimensions at finite temperature. We have shown that the zero
temperature part of the parity violating quartic action is unique and,
in fact, so is the structure of the complete parity violating part of
the  effective
action. In evaluating the temperature dependent contribution, we have
pointed out various obstacles that one has to face and have
systematically shown how one can handle these in a given
calculation. Of importance is the non-analyticity of thermal
amplitudes as well as of the thermal effective actions. We have discussed
this in detail for the CS term (self-energy) as well as for the parity
violating part of the four photon amplitude. In particular, we have
shown that the behavior of the leading amplitudes and, therefore, the
leading effective actions in the long wave and static limits are quite
distinct at high temperature. Furthermore, while the leading term in
the quartic effective action is extensive (but non-local) in the long
wave limit, it is non-extensive in the static limit. We have found
that, in  the long wave limit, {\it large} gauge invariance is
manifest order by order. In contrast, it appears to be violated order
by order in the static limit.

These results can be understood intuitively from the following
heuristic arguments. Note that, in $2+1$ dimensions, we can always
write
\beN{67a}
A_{0} (t,\vec{x}) =  {1\over \beta}\int_{0}^{\beta}
dt'\,A_{0}(t',\vec{x}) + \partial_{t}\Omega (t,\vec{x}),
\ee
\beN{67}
A_{i} (t,\vec{x}) = {1\over \nabla^{2}}\,\partial^{j}F_{ij}
(0,\vec{x}) + \left(\int_{0}^{t} - {t\over
\beta}\int_{0}^{\beta}\right)dt'\,E_{i}(t',\vec{x}) +
\partial_{i}\Omega(t,\vec{x}),
\ee
where,
\beN{68}
\Omega(t,\vec{x}) = \left(\int_{0}^{t} - {t\over
\beta}\int_{0}^{\beta}\right) dt'\,A_{0}(t',\vec{x}) - 
{1\over\nabla^{2}}\,\vec{\nabla}\cdot \vec{A}(0,\vec{x})
\ee
Thus, in a particular gauge, we can think of $a_0(\vec{x})$,
$B(0,\vec{x})$ and $\vec{E}(t,\vec{x})$ as representing the physically
meaningful variables. From this, it is clear that, in the long wave
limit, the only meaningful variable is the electric field which is
both {\it small} and {\it large} gauge invariant. Consequently, the
effective action, in this limit, would be {\it large} gauge invariant
order by order. In contrast, in the static limit, all of the three
variables are meaningful implying that the leading term (in
derivatives) would involve an odd number of $a_0(\vec{x})$ and a single
$B(0,\vec{x})$. Of course, there can be other terms, but they will be
higher order in the number of derivatives. Furthermore, order by
order, the leading term would violate {\it large} gauge invariance.

We have written down a {\it large} gauge Ward identity that the
leading order terms of the parity violating effective action in the
static  limit must satisfy for {\it large} gauge invariance to
hold. This identity can be solved to obtain the leading, all order parity
violating effective action which coincides with the action proposed
earlier in a restrictive gauge background. However, it is worth
remembering that this does not represent the full effective action --
rather, it only represents the leading term of the full parity
violating effective action.

This study has been carried out within the context of an Abelian gauge
theory as a first step towards understanding the question of 
{\it large} gauge invariance at finite temperature. The main interest is,
of course, the study of this issue within the context of a
non-Abelian gauge theory, which is work in progress.

This work was supported in part by U.S. Dept. Energy Grant DE-FG
02-91ER40685, NSF-INT-9602559 as well as by CNPq, Brazil.

\end{document}